\documentstyle[prd,aps,amsmath,latexsym,amssymb,floats]{revtex}
\def \slas{\kern -6.2pt /}
\def \sla{\kern -5.4pt /}
\def \sl{\kern -4.0pt /}
\def \Cslas{\kern -6.8pt /}
\def \Dslas{\kern -7.4pt /}
\def \slass{\kern -7.4pt /}

\def \ii{{\mathrm{i}}}
\def \d{{\mathrm{d}}}

\def \pd{\partial}
\def \e{{\mathrm{e}}}

\def \lcx{\tilde{x}}
\def \tl#1{\overset{\kern 2pt\circ}{#1}}
\def \tll#1{\overset{\kern -1pt\circ}{#1}}
\def \TL#1{\overset{\kern -28pt \circ}{#1}}
\def \TLL#1{\overset{\kern -7pt \circ}{#1}}

\begin{document}

\title{$\rho$-Meson wave functions from nonlocal light-cone operators 
with definite twist}
\author{Markus Lazar\thanks{E-mail: lazar@itp.uni-leipzig.de}}
\address{Center for Theoretical Studies
	 and Institute of Theoretical Physics,\\
	Leipzig University, 
	Augustusplatz~10, D-04109~Leipzig, Germany} 
\date{\today}    
	
\maketitle
\hspace*{1cm}
\begin{abstract}
We introduce chiral-even and chiral-odd meson wave functions
as vacuum-to-meson matrix elements of bilocal quark operators
with well-defined (geometric) twist. Thereby, we achieve
a Lorentz invariant classification of these distributions
which differ from the conventional ones by explicitly taking
into account the 
 trace terms. The relations between conventional and new 
wave functions are given. 
\end{abstract}

\section{Introduction}
Recently, we have introduced a group theoretical procedure to 
decompose local and nonlocal operators~\cite{glr99b,gl00a}, which are 
important in different hard scattering processes, into operators of
definite twist.  This procedure is based on the notion of 
{\it geometric} twist = mass dimension -- (Lorentz) spin, $\tau=d-j$, 
originally introduced by Gross and Treiman~\cite{Gross}. 
We are interested on nonlocal light-cone~(LC) operators 
and their matrix elements which describe different phenomenological distribution
amplitudes, e.g., parton distribution functions  as well as hadronic wave 
functions. 
The classification of these functions suffers from the fact, that one and 
the same operator by its twist decomposition contributes to different 
wave functions. It is thus necessary to disentangle these various 
contributions of the original operator and to specify the contributions
of the trace terms. 

Jaffe and Ji~\cite{Jaf92} proposed the notion of {\it dynamical} 
twist ($t$) by counting powers $Q^{2-t}$ which is directly related to 
the power by which the corresponding distributions contribute to the 
scattering amplitudes. They classified the parton distribution functions 
which correspond to the independent tensor structure of the matrix 
elements of bilocal quark-antiquark operators.
Recently, Ball {\it et al.} have used this pattern for the 
classification of leading and higher twist wave
functions of $\rho$-vector mesons in QCD~\cite{ball96,ball98,ball99}. 
They found eight independent (two-particle) meson wave functions.
One key ingredient in their approach was the use of QCD equation of motion
in order to obtain integral representations for wave 
functions that are not dynamically independent.
Ball {\it et al.} have already pointed out that the 
geometric twist is more convenient to discuss higher twist effects on a 
reliable basis~\cite{ball98}.
On the other hand, the pion wave functions are earlier investigated  
in similar way~\cite{braun89,braun90}.
Meson wave functions and form factors have been discussed in the 
framework of local operator product expansion and in the infinite 
momentum frame by~\cite{brodsky79,brodsky80,rady80,chern84,shifman,craigie}
and in the framework of nonlocal operator product expansion~\cite{geyer85,geyer87,mul94}.

However, such different notion of dynamical twist is only defined for 
the matrix elements of operators, it is not Lorentz invariant and also 
not simply related to the contributions of (higher) geometric twist.
Quit recently, in order to show the relation between the different 
definitions of twist, we have calculated the forward quark distribution 
functions by the help of our bilocal quark operators with definite 
twist~\cite{gl00d}.
We proved that the two definitions of twist do not coincide at higher 
orders and we gave relation between Jaffe and Ji's distribution function 
and our distributions with geometrical twist.
 
The aim of this paper is to present the (two-particle) {\em meson wave 
functions} which are related to the nonlocal LC-operators of different
geometric twist. In that framework, it is possible to investigate 
in an unique manner the contributions resulting from the traces of the 
operators having well-defined twist.

\section{Twist decomposition of bilinear quark operators}
In~\cite{glr99b,gl00a} we explained a procedure of twist decomposition 
for the  bilocal
quark operators on the light-cone 
which are relevant for the meson wave functions, too:
\begin{align}
\label{O_ent}
O_{\alpha}(\kappa_1\lcx,\kappa_2\lcx)&=
\bar{u}(\kappa_1\lcx)
\gamma_{\alpha}
U(\kappa_1\lcx,\kappa_2\lcx)d(\kappa_2\lcx),\\
\label{M_ent}
M_{[\alpha\beta]}(\kappa_1\lcx,\kappa_2\lcx)
&=
\bar{u}(\kappa_1\lcx)
\sigma_{\alpha\beta}
U(\kappa_1\lcx,\kappa_2\lcx)d(\kappa_2\lcx),
\end{align}
with the path ordered gauge factor along the straight line connecting the 
points $\kappa_1\lcx$ and $\kappa_2\lcx$:
\begin{equation}
\label{phase}
U(\kappa_1 \lcx, \kappa_2 \lcx)
 =  P \exp\left\{\ii g
\int_{\kappa_1}^{\kappa_2} \d w \,\lcx^\mu A_\mu(w \lcx)
\right\}.
\end{equation}
The resulting decomposition for the vector and skew tensor operators is:
\begin{align}
\label{O_tw_nl}
O_{\alpha}(\kappa_1\lcx,\kappa_2\lcx)&=
 O^{\mathrm{tw2}}_{\alpha}(\kappa_1\lcx,\kappa_2\lcx)
+O^{\mathrm{tw3}}_{\alpha}(\kappa_1\lcx,\kappa_2\lcx)
+O^{\mathrm{tw4}}_{\alpha}(\kappa_1\lcx,\kappa_2\lcx)
,\\
\label{M_tw_nl}
M_{[\alpha\beta]}(\kappa_1\lcx,\kappa_2\lcx)&=
M^{\mathrm{tw2}}_{[\alpha\beta]}(\kappa_1\lcx,\kappa_2\lcx)
+M^{\mathrm{tw3}}_{[\alpha\beta]}(\kappa_1\lcx,\kappa_2\lcx)
+M^{\mathrm{tw4}}_{[\alpha\beta]}(\kappa_1\lcx,\kappa_2\lcx)
\end{align}
with
\begin{table*}[h]
\begin{align}
\label{O2sca}
O^{\rm tw2}(\kappa_1\lcx,\kappa_2\lcx)
&\equiv
\lcx^\mu O_\mu(\kappa_1\lcx,\kappa_2\lcx)
=
\bar{u}(\kappa_1\lcx)(\gamma\lcx)
U(\kappa_1\lcx, \kappa_2\lcx) d(\kappa_2\lcx)
\\
\label{O2vec}
O^{\mathrm{tw2}}_{\alpha}(\kappa_1\lcx,\kappa_2\lcx)
&=
\int_{0}^{1} \d t
\Big(\pd_\alpha +
\hbox{\Large$\frac{1}{2}$}(\ln t)\,x_\alpha\square\Big)
x^\mu
O_\mu(\kappa_1 t x, \kappa_2 t x)
\big|_{x=\tilde{x}}
\\
\label{O3vec}
O^{\mathrm{tw3}}_{\alpha}
(\kappa_1\lcx,\kappa_2\lcx)
&=
\int_{0}^{1}\d t 
\Big(\delta_\alpha^\mu(x\pd)-
x^\mu\pd_\alpha-(1+2\ln t )x_\alpha\pd^\mu
-(\ln t )\, x_\alpha x^\mu\square
\Big) O_\mu(\kappa_1 t  x, \kappa_2 t  x)
\big|_{x=\tilde{x}}
\\
\label{O4vec}
O^{\mathrm{tw4}}_{\alpha}
(\kappa_1\lcx,\kappa_2\lcx)
&=\lcx_\alpha
\int_{0}^{1}\d t \Big(
(1+\ln t )\pd^\mu+
\hbox{\Large$\frac{1}{2}$}(\ln t )\,x^\mu \square
\Big)
O_\mu(\kappa_1 t  x, \kappa_2 t x)\big|_{x=\tilde{x}}
\\
\label{M_tw2_ten}
M^{\mathrm{tw2}}_{[\alpha\beta]}(\kappa_1\lcx,\kappa_2\lcx)
&=
 \int_{0}^{1}\d t \Big\{2t \,
\pd_{[\beta}\delta_{\alpha]}^\mu
-(1-t )\big(2x_{[\alpha}\pd_{\beta]}\pd^\mu
-x_{[\alpha}\delta_{\beta]}^\mu\square\big)\Big\}
x^\nu M_{[\mu\nu]}(\kappa_1 t x, \kappa_2 t x)
\big|_{x=\tilde{x}}
\\
\label{M_tw3_ten}
M^{\rm tw3}_{[\alpha\beta]}(\kappa_1\lcx,\kappa_2\lcx)
&=
\int_{0}^{1}\d t 
\Big\{t \big((x\pd)\delta_{[\beta}^\nu
- 2x^\nu\pd_{[\beta}\big)\delta_{\alpha]}^\mu
+ \hbox{\Large$\frac{1-t ^2}{t }$}
\Big(x_{[\alpha}\big(\delta_{\beta]}^{[\mu} (x\pd)-
x^{[\mu}\pd_{\beta]}\big)\pd^{\nu]}
-x_{[\alpha}\delta_{\beta]}^{[\mu}x^{\nu]}\square
\nonumber
\\
&\qquad
-x_{[\alpha}\pd_{\beta]}x^{[\mu}\pd^{\nu]}
\Big)\Big\}
M_{[\mu\nu]}(\kappa_1 t x, \kappa_2 t x)\big|_{x=\lcx}
\\
\label{M_tw4_ten}
M^{\rm tw4}_{[\alpha\beta]}(\kappa_1\lcx,\kappa_2\lcx)
&=
\int_{0}^{1}{\d t }
\hbox{\Large$\frac{1-t }{t }$}
\Big\{x_{[\alpha}\delta_{\beta]}^{[\mu}
x^{\nu]}\square
-2x_{[\alpha}\big(
\delta_{\beta]}^{[\mu} (x\pd)
-
x^{[\mu}\pd_{\beta]}
\big)\pd^{\nu]}
\Big\}
M_{[\mu\nu]}(\kappa_1 t x, \kappa_2 t x)\big|_{x=\lcx} .
\end{align}
\end{table*}

Let us remark that the vector and skew tensor operators of  
twist $\tau$, $O^{(\tau)}_\alpha(\kappa_1\lcx, \kappa_2 \lcx)$ and
$M^{(\tau)}_{[\alpha\beta]}(\kappa_1\lcx, \kappa_2 \lcx)$,
are obtained from the original (undecomposed)
operators $O_\alpha(\kappa_1\lcx, \kappa_2 \lcx)$ and 
$M_{[\alpha\beta]}(\kappa_1\lcx, \kappa_2 \lcx)$, 
Eqs.~(\ref{O_ent}) and (\ref{M_ent}),
by the application of the corresponding twist 
projectors (including the $ t $--integrations),
${\cal P}^{(\tau)\mu}_\alpha$ and
${\cal P}^{(\tau)[\mu\nu]}_{[\alpha\beta]}$, 
defined by Eqs.~(\ref{O2vec}) -- (\ref{O4vec})
and (\ref{M_tw2_ten}) -- (\ref{M_tw4_ten}), respectively:
\begin{align}
\label{OPROJ}
O^{(\tau)}_\alpha(\kappa_1\lcx, \kappa_2 \lcx) &= 
({\cal P}^{(\tau)\mu}_\alpha O_\mu)(\kappa_1\lcx, \kappa_2 \lcx)
\\
\label{MPROJ}
M^{(\tau)}_{[\alpha\beta]}(\kappa_1\lcx, \kappa_2 \lcx) &= 
({\cal P}^{(\tau)[\mu\nu]}_{[\alpha\beta]} 
M_{[\mu\nu]})(\kappa_1\lcx, \kappa_2 \lcx)
\\
\intertext{with}
\label{OProj}
({\cal P}^{(\tau)}\times{\cal P}^{(\tau')})^{\mu}_\alpha
&=
\delta^{\tau \tau'}
{\cal P}^{(\tau)\mu}_\alpha .
\\
\label{MProj}
({\cal P}^{(\tau)}\times{\cal P}^{(\tau')})^{[\mu\nu]}_{[\alpha\beta]}
&=
\delta^{\tau \tau'}
{\cal P}^{(\tau)[\mu\nu]}_{[\alpha\beta]} .
\end{align}
These projection operators contain the corresponding symmetry of 
the Young pattern. 

Additionally, the twist--2 vector operator, Eq.~(\ref{O2vec}), is related to the
corresponding scalar operator, (\ref{O2sca}). This leads to relations of the corresponding
meson wave functions, Eqs.~(\ref{matrix_O_t2_sca}) and 
(\ref{matrix_O_t2_nl}) below.

\section{The $\rho$-meson matrix elements of LC--operators with twist}
In this section we define the meson wave functions 
for the bilinear LC-quark operators with definite twist sandwiched between the 
vacuum and the meson state. As usual,
the matrix elements of the meson are related to the 
meson momentum $P_\mu$ and meson polarization vector $e^{(\lambda)}_\mu$, respectively,
with $P^2=m_\rho^2$, $e^{(\lambda)}\cdot e^{(\lambda)}=-1$, $P\cdot e^{(\lambda)}=0$, 
$m_\rho$ denoting the meson mass.

Taking meson matrix elements of Eqs.~(\ref{O2sca}) -- (\ref{M_tw4_ten}) 
we see that, observing the correct tensor structure by the use 
of $P_\mu, e^{(\lambda)}_\mu$ and $\lcx_\mu$, we may introduce any parametrization 
for the matrix elements of the undecomposed operators, 
e.g.,
\begin{align}
\label{FULL}
\langle 0|O_\alpha(\kappa_1\lcx,\kappa_2\lcx)|\rho(P,\lambda)\rangle
=f_\rho m_\rho\int_0^1\d u \Big( e^{(\lambda)}_\alpha \widetilde \Phi_1 (u,\mu^2) 
+P_\alpha\frac{(e^{(\lambda)}\lcx)}{\lcx P} \widetilde \Phi_2(u,\mu^2)
+\lcx_\alpha \frac{m_\rho^2(e^{(\lambda)}\lcx)}{(\lcx P)^2} \widetilde \Phi_3(u,\mu^2) \Big)
{\e^{\ii\kappa\xi(\lcx P)}},
\end{align}
where $\mu^2$ denotes the renormalization scale and $f_\rho$ is the vector
meson decay constant. We defined $\kappa=(\kappa_1-\kappa_2$)/2. 
According to the above projection properties we are able 
to introduce one (and only one) wave function for any operator of 
definite twist. 

We start with the chiral-even {\em scalar operator}.
The  matrix element of this nonlocal twist-2 operator, 
Eq.~(\ref{O2sca}), taken between the vacuum $\langle 0|$ and the meson 
state $|\rho(P,\lambda)\rangle$ reads
\begin{align}
\label{matrix_O_t2_sca}
\langle 0|O^{\text{tw2}}(\kappa_1\lcx,\kappa_2\lcx)|\rho(P,\lambda)\rangle
=
f_\rho m_\rho(e^{(\lambda)}\lcx)\int_0^1\d u\,\Phi^{(2)}(u,\mu^2)\,{\e^{\ii\kappa \xi(\lcx P)}}
=f_\rho m_\rho(e^{(\lambda)}\lcx)\sum_{n=0}^\infty \frac{(\ii\kappa(\lcx P))^n}{n!}\,\Phi^{(2)}_n(\mu^2) .
\end{align}
Here, $\Phi^{(2)}(u,\mu^2)$ is the twist-2 meson 
wave function which is related to the corresponding moments 
\begin{align}
\Phi^{(2)}_n(\mu^2)=\int_0^1\d u\, \xi^n \Phi^{(2)}(u,\mu^2).
\nonumber
\end{align} 
For brevity, we use the shorthand notation
\begin{align}
\xi=u-(1-u)=2u-1.
\end{align}
The wave functions describe the probability amplitudes to find the 
$\rho$-meson in a state with the minimal number of constituents, e.g., 
quark which carries momentum fractions $u$ and antiquark with
$1-u$, respectively.

Now we consider the chiral-even {\em  vector operator}.
Using the projection properties (\ref{OPROJ}), together
with (\ref{OProj}), we introduce the  meson wave functions 
$\Phi^{(\tau)}(u, \mu^2)$ of twist $\tau$ by
\begin{align}
\label{Gfct}
\langle 0|O^{(\tau)}_\alpha(\kappa_1\lcx, \kappa_2 \lcx)|\rho(P,\lambda)\rangle 
&\equiv 
\langle 0|({\cal P}^{(\tau)\beta}_\alpha 
O^{(\tau)}_\beta)(\kappa_1\lcx, \kappa_2 \lcx)|\rho(P,\lambda)\rangle 
= f_\rho m_\rho{\cal P}^{(\tau)\beta}_\alpha
\Big(e^{(\lambda)}_\beta\int_0^1\d u\, \Phi^{(\tau)}(u,\mu^2)\,
{\e^{\ii\kappa\xi(\lcx P)}}\Big)\ ,
\end{align} 
which, for $\tau =2$,  is  consistent with (\ref{matrix_O_t2_sca}).
Using the twist projection operators as they are determined by Eqs.~(\ref{O2vec})
-- (\ref{O4vec}) we obtain for the twist-2 operator (from now on we suppress $\mu^2$)
\begin{align}
\label{matrix_O_t2_nl}
&\langle 0|O^{\text{tw2}}_{\alpha}(\kappa_1\lcx,\kappa_2\lcx)|\rho(P,\lambda)\rangle
=f_\rho m_\rho\int_0^1\d t \Big[\pd_\alpha+\frac{1}{2} \big(\ln t\big) 
x_\alpha\square\Big]
(e^{(\lambda)}x)\int_0^1\d u\, \Phi^{(2)}(u)\,{\e^{\ii\kappa  \xi t(xP)}}
\big|_{x=\tilde{x}}\nonumber\\
&\qquad\qquad=f_\rho m_\rho\int_0^1\d t \int_0^1\d u\,\Phi^{(2)}(u)
\Big[e^{(\lambda)}_\alpha+\ii\kappa \xi t
P_\alpha(e^{(\lambda)}\lcx)
+\frac{\tilde{x}_\alpha}{2}(e^{(\lambda)}\lcx)m_\rho^2
(\ii\kappa \xi t)^2\big(\ln t \big)
\Big]
\e^{\ii\kappa  \xi t(\tilde{x}P)}\\
&\qquad\qquad=f_\rho m_\rho
\sum_{n=0}^\infty \frac{(\ii\kappa(\lcx P))^n}{n!}\Phi^{(2)}_n\Big\{
\frac{1}{n+1}\Big( e^{(\lambda)}_\alpha+n P_\alpha \frac{e^{(\lambda)}\lcx}
{\lcx P}\Big)
-\frac{n(n-1)}{2(n+1)^2}\lcx_\alpha\frac{m_\rho^2 (e^{(\lambda)}\lcx)}{(\lcx P)^2}\Big\},
\label{matrix_O_t2_l}
\end{align}
and for the higher twist operators
\begin{align}
\label{matrix_O_t3_nl}
&\langle 0|O^{\mathrm{tw3}}_{\alpha}(\kappa_1\lcx,\kappa_2\lcx)|\rho(P,\lambda)\rangle
=f_\rho m_\rho \int_0^1\!\!\d t \int_0^1\!\!\d u\, \Phi^{(3)}(u)
\Big[\Big(e^{(\lambda)}_\alpha (\tilde{x}P)
-P_\alpha (e^{(\lambda)}\lcx)\Big)\ii\kappa \xi t
-\tilde{x}_\alpha m_\rho^2(e^{(\lambda)}\lcx)
(\ii\kappa \xi t)^2\ln t \Big]
\e^{\ii\kappa \xi t(\tilde{x}P)}\\
&\qquad\qquad=f_\rho m_\rho\sum_{n=1}^\infty \frac{(\ii\kappa(\lcx P))^n}{n!}\Phi^{(3)}_n\Big\{
\frac{n}{n+1}\Big( e^{(\lambda)}_\alpha-P_\alpha \frac{e^{(\lambda)}\lcx}{\lcx P}\Big)
+\frac{n(n-1)}{(n+1)^2}\lcx_\alpha\frac{m_\rho^2 (e^{(\lambda)}\lcx)}{(\lcx P)^2}\Big\},
\\
\label{matrix_O_t4_nl}
&\langle 0|O^{\mathrm{tw4}}_{\alpha}(\kappa_1\lcx,\kappa_2\lcx)|\rho(P,\lambda)\rangle
=
\frac{1}{2}f_\rho m_\rho
\lcx_\alpha(e^{(\lambda)}\lcx)m_\rho^2
\int_0^1\d t \int_0^1\d u\, \Phi^{(4)}(u)
(\ii\kappa \xi t)^2(\ln t) \,\e^{\ii\kappa \xi t(\tilde{x}P)}\\
&\qquad\qquad=-f_\rho m_\rho \sum_{n=2}^\infty \frac{(\ii\kappa(\lcx P))^n}{n!}\Phi^{(4)}_n
\frac{n(n-1)}{2(n+1)^2}\lcx_\alpha\frac{m_\rho^2 (e^{(\lambda)}\lcx)}{(\lcx P)^2}.
\end{align}

In the first line of any equation we have given the nonlocal matrix element
of definite twist, 
and in the second line, after expanding the nonlocal expression, we introduced the 
moments of the meson wave functions; 
thereby, the $ t $--integrations contribute the additional $n$-dependent factors. 
Obviously, the trace terms which have been explicitly subtracted  
are proportional to $m_\rho^2$. According to the terminology of
Jaffe and Ji, they contribute to {\it dynamical} twist-4. 
For the twist--2 operator we observe that the terms proportional to 
$e^{(\lambda)}_\alpha$, $P_\alpha$ and $\lcx_\alpha$ have contributions starting 
with the zeroth, first and second moment, respectively.
The twist--3 operator starts with the first moment, and the
twist--4 operator starts with the second moment. Analogous statements
also hold for the twist--$\tau$ operators below.
\\
Putting together the different twist contributions we obtain,
after replacing $\ii\kappa \xi t(\lcx P)$ by $ t \pd/\pd t $
and performing partial integrations, the following 
matrix element of the original operator ($\zeta=\kappa(\lcx P)$)
\begin{align}
\label{O_full}
\langle 0|O_{\alpha}(\kappa_1\lcx,\kappa_2\lcx)|\rho(P,\lambda)\rangle
&=f_\rho m_\rho\Bigg[
P_\alpha\frac{e^{(\lambda)}\lcx }{\lcx P}\int_0^1\d u
\Big(\Phi^{(2)}(u)-\Phi^{(3)}(u)\Big)[e_0(\ii\zeta\xi)-e_1(\ii\zeta\xi)]\\
&+e^{(\lambda)}_\alpha\int_0^1\d u
\Big(\Phi^{(2)}(u)e_1(\ii\zeta\xi)+\Phi^{(3)}(u)
[e_0(\ii\zeta\xi)-e_1(\ii\zeta\xi)]\Big)\nonumber\\
&-\frac{1}{2}\lcx_\alpha
\frac{e^{(\lambda)}\lcx }{(\lcx P)^2}m_\rho^2\int_0^1\!\!\d u
\Big(\Phi^{(2)}(u)-2\Phi^{(3)}(u)+\Phi^{(4)}(u)\Big)
\Big[e_0(\ii\zeta\xi)-3e_1(\ii\zeta\xi)
+2\int_0^1\!\!\d t\,  e_1(\ii\zeta \xi t)\Big]
\Bigg], \nonumber
\end{align}
where we used the following ``truncated exponentials''
\begin{align}
e_0(\ii\zeta\xi)=\e^{\ii\zeta\xi},\qquad 
e_1(\ii\zeta\xi)=\int_0^1\!\!\d t \,\e^{\ii\zeta  \xi t}
=\frac{\e^{\ii\zeta\xi}-1}{\ii\zeta\xi},
\quad\cdots\quad,
e_{n+1}(\ii\zeta\xi)
=\frac{(-1)^{n}}{n!}\int_0^1\!\!\d t \, t ^n\,\e^{\ii\zeta \xi t}.
\end{align}
As it should be the application of the projection operators
${\cal P}^{(\tau)\beta}_\alpha$ onto (\ref{O_full}) reproduces
the matrix elements (\ref{matrix_O_t2_nl}) --
(\ref{matrix_O_t4_nl}). In comparison with (\ref{FULL})
we also observe that the wave functions are accompanied
not simply by the exponentials, $e_0(\ii\zeta \xi)$, but by more 
involved combinations whose series expansion directly leads to the
representations with the help of moments.
Let us mention that the twist-2 part of Eq.~(\ref{O_full}) and the local 
expression (\ref{matrix_O_t2_l}) are in agreement with 
Stoll's result~\cite{stoll} for the light-cone twist-2 operator.

Now we consider the chiral-even {\em vector operator}
$O_{5\alpha}(\kappa_1\lcx,\kappa_2\lcx)=
\bar{u}(\kappa_1\lcx)\gamma_{\alpha}\gamma_5
U(\kappa_1\lcx,\kappa_2\lcx)d(\kappa_2\lcx)$, which
obeys relations Eqs.~(\ref{O2sca}) -- (\ref{O4vec}), as well as 
(\ref{OPROJ}) and (\ref{OProj}) with the {\em same} projection operator
as the vector operator. Let us introduce the corresponding
meson wave functions $\Xi^{(\tau)}(z)$ of twist $\tau$ by
\begin{align}
\label{Ffct}
\langle 0|O^{(\tau)}_{5\alpha}(\kappa_1\lcx, \kappa_2 \lcx)|\rho(P,\lambda)\rangle 
= \frac{1}{2}\Big(f_\rho -f_\rho^{\rm T}\frac{m_u+m_d}{m_\rho}\Big)m_\rho
{\cal P}^{(\tau)\beta}_\alpha
\Big(\epsilon_\beta^{\ \,\gamma\mu\nu}
e^{(\lambda)}_\gamma P_\mu\lcx_\nu\int_0^1\d u\, \Xi^{(\tau)}(u)\,
{\e^{\ii\kappa\xi(\lcx P)}}\Big)\ ,
\end{align} 
where $f^{\rm T}_\rho$ denotes the tensor decay constant.
The vacuum-to-meson matrix elements of these {\em vector operators} of 
twist $\tau$ are obtained as follows:
\begin{align}
\label{matrix_O_tw3_nl}
\langle 0|
O^{\text{tw3}}_{5\alpha}(\kappa_1\lcx,\kappa_2\lcx)|\rho(P,\lambda)\rangle
&=
\frac{1}{2}\Big(f_\rho -f_\rho^{\rm T}\frac{m_u+m_d}{m_\rho}\Big)m_\rho
\epsilon_\alpha^{\ \,\beta\mu\nu}
e^{(\lambda)}_\beta P_\mu \lcx_\nu
\int_0^1\!\!\d t \int_0^1\!\!\d u\, \Xi^{(3)}(u)
\big(1+\ii\kappa \xi t(\lcx P)\big)
\e^{\ii\kappa \xi t(\tilde{x}P)}\\
&=\frac{1}{2}\Big(f_\rho -f_\rho^{\rm T}\frac{m_u+m_d}{m_\rho}\Big)m_\rho
\epsilon_\alpha^{\ \,\beta\mu\nu}
e^{(\lambda)}_\beta P_\mu\lcx_\nu
\sum_{n=0}^\infty \frac{(\ii\kappa(\lcx P))^n}{n!}\,\Xi^{(3)}_n .
\end{align}
We see that only the twist-3 operator give a contribution and the twist-2 
and twist-4 operator as well as 
all trace terms vanish. 
Thus, the matrix element of the original operator reads
\begin{align}
\label{matrix_O_nl}
&\langle 0|O^{}_{5\alpha}(\kappa_1\lcx,\kappa_2\lcx)|\rho(P,\lambda)\rangle
=\frac{1}{2}\Big(f_\rho -f_\rho^{\rm T}\frac{m_u+m_d}{m_\rho}\Big)m_\rho
\epsilon_\alpha^{\ \,\beta\mu\nu}
e^{(\lambda)}_\beta P_\mu\lcx_\nu
\int_0^1\d u\, \Xi^{(3)}(u) e_0(\ii\zeta \xi).
\end{align}

The matrix element of the simplest bilocal {\em scalar operator} 
arises as
\begin{align}\label{}
\langle 0|\bar{u}(\kappa_1\lcx)
U(\kappa_1\lcx,\kappa_2\lcx)d(\kappa_2\lcx)|\rho(P,\lambda)\rangle
&=-\ii\Big( f^{\rm T}_\rho-f_\rho \frac{m_u+m_d}{m_\rho}\Big)
\big(e^{(\lambda)}\lcx\big)m_\rho^2
\int_0^1\d u\, \Upsilon^{(3)}(u){\e^{\ii\kappa\xi(\lcx P)}}
\\
&=-\ii\Big( f^{\rm T}_\rho-f_\rho \frac{m_u+m_d}{m_\rho}\Big)
\big(e^{(\lambda)}\lcx\big)m_\rho^2
\sum_{n=0}^\infty \frac{(\ii\kappa(\lcx P))^n}{n!}\,\Upsilon^{(3)}_n ,
\end{align}
where $\Upsilon^{(3)}(u)$ is another spin-independent twist-3 meson wave 
function.

Now, we consider the matrix elements of the chiral-odd {\em  skew tensor 
operators}.
The corresponding wave functions are introduced by
\begin{align}
\label{Hfct}
\langle 0|M^{(\tau)}_{\alpha\beta}(\kappa_1\lcx, \kappa_2 \lcx)|\rho(P,\lambda)\rangle 
&=\ii f^{\rm T}_\rho {\cal P}^{(\tau)[\mu\nu]}_{[\alpha\beta]}
\Big((e^{(\lambda)}_\mu P_\nu-e^{(\lambda)}_\nu P_\mu) 
\int_0^1\d u\, \Psi^{(\tau)}(u)\,{\e^{\ii\kappa\xi(\lcx P)}}\Big).
\end{align}
The matrix elements of the skew tensor operators of twist $\tau$ 
are obtained using the projectors determined by Eqs.~(\ref{M_tw2_ten})
-- (\ref{M_tw4_ten}):
\begin{align}
\label{matrix_M_tw2_nl}
&\langle 0|
M^{\mathrm{tw2}}_{[\alpha\beta]}(\kappa_1\lcx,\kappa_2\lcx)|\rho(P,\lambda)\rangle
=\ii f_\rho^{\rm T}\int_0^1\d t \int_0^1\d u\, \Psi^{(2)}(u)
\Big[2 t \, e^{(\lambda)}_{[\alpha} P_{\beta]}\big(2+\ii\kappa \xi t(\lcx P)\big)
\nonumber\\
&\qquad\qquad\qquad\qquad\qquad\qquad
+\left(1- t \right)m_\rho^2
\lcx_{[\alpha}\Big\{4(\ii\kappa \xi t)e^{(\lambda)}_{\beta]}
+(\ii\kappa \xi t)^2\big(e^{(\lambda)}_{\beta]}(\lcx P)
+P_{\beta]}(e^{(\lambda)}\lcx)\big)\Big\}
\Big]\e^{\ii\kappa \xi t(\lcx P)}
\\
&\quad
= \ii f_\rho^{\rm T}\sum_{n=0}^\infty \frac{(\ii\kappa(\lcx P))^n}{n!}\Psi^{(2)}_n\Big\{
 2e^{(\lambda)}_{[\alpha}P_{\beta]}
+\frac{4n}{(n+2)(n+1)}\frac{m_\rho^2}{\lcx P}\lcx_{[\alpha}e^{(\lambda)}_{\beta]}
+\frac{n(n-1)}{(n+2)(n+1)}\frac{m_\rho^2}{\lcx P}\lcx_{[\alpha}\Big(e^{(\lambda)}_{\beta]}
+P_{\beta]}\frac{e^{(\lambda)}\lcx}{\lcx P}\Big)\Big\},
\\
&\langle 0|
M^{\mathrm{tw3}}_{[\alpha\beta]}(\kappa_1\lcx,\kappa_2\lcx)|\rho(P,\lambda)\rangle
=\ii f_\rho^{\rm T}\int_0^1\d t \, \frac{1- t ^2}{ t }
\int_0^1\d u\,\ii\kappa \xi t\, \Psi^{(3)}(u) m_\rho^2
\lcx_{[\alpha}\Big\{ e^{(\lambda)}_{\beta]}
+\ii\kappa \xi t P_{\beta]}(e^{(\lambda)}\lcx)\Big\}
\e^{\ii\kappa \xi t(\lcx P)}\\
&\phantom{\langle 0|
M^{\mathrm{tw2}}_{\alpha\beta}(\kappa_1\lcx,\kappa_2\lcx)|\rho(P,\lambda)\rangle}
=-\ii f_\rho^{\rm T}\sum_{n=1}^\infty \frac{(\ii\kappa(\lcx P))^n}{n!}\Psi^{(3)}_n
\frac{2}{n+2}\frac{m_\rho^2}{\lcx P}\lcx_{[\alpha}\Big(e^{(\lambda)}_{\beta]}
+(n-1)P_{\beta]}\frac{e^{(\lambda)}\lcx}{\lcx P}\Big),
\\
\label{matrix_M_tw4_nl}
&\langle 0|
M^{\mathrm{tw4}}_{[\alpha\beta]}(\kappa_1\lcx,\kappa_2\lcx)|\rho(P,\lambda)\rangle
=-\ii f_\rho^{\rm T}\int_0^1\d t \,\frac{1- t }{ t }
\int_0^1\d u\, (\ii  t \kappa\xi )^2 \Psi^{(4)}(u)m_\rho^2
\lcx_{[\alpha}\Big\{e^{(\lambda)}_{\beta]}(\lcx P)-P_{\beta]}(e^{(\lambda)}\lcx)\Big\}
\e^{\ii\kappa \xi t(\lcx P)}\\
&\phantom{\langle 0|
M^{\mathrm{tw2}}_{\alpha\beta}(\kappa_1\lcx,\kappa_2\lcx)|\rho(P,\lambda)\rangle}
=-\ii f_\rho^{\rm T}\sum_{n=2}^\infty \frac{(\ii\kappa(\lcx P))^n}{n!}\Psi^{(4)}_n
\frac{n-1}{n+1}\frac{m_\rho^2}{\lcx P}\lcx_{[\alpha}\Big(e^{(\lambda)}_{\beta]}
-P_{\beta]}\frac{e^{(\lambda)}\lcx}{\lcx P}\Big).
\end{align}

Again, the moments of wave functions of twist $\tau = 2,3$
and $4$ start with $n=0,1$ and $2$, respectively. In addition,
we remark that only those terms of the operator (\ref{M_tw3_ten}) 
contribute to the twist--3 wave function which result from
the trace terms of (\ref{M_tw2_ten}). Analogous to the vector case
the forward matrix element of the `true' twist--3 part of (\ref{M_tw3_ten})
vanishes. 
In Eq.~(\ref{matrix_M_tw4_nl}) 
only the twist--4 operator contributes which result from
the trace terms of (\ref{M_tw2_ten}).
After multiplication of (\ref{matrix_M_tw4_nl}) with
$\lcx_\alpha$ (or $\lcx_\beta$) the matrix element vanishes because the 
corresponding vector operator does not contain any twist-four contribution.
 
The matrix element of the original skew tensor operator is obtained as
\begin{align}
\label{M_full_JJ}
\langle 0|M_{[\alpha\beta]}(\kappa_1\lcx,\kappa_2\lcx)|\rho(P,\lambda)\rangle
&=
\ii f_\rho^{\rm T}\Big[
2e^{(\lambda)}_{[\alpha}P_{\beta]}
\int_0^1\d u\, \Psi^{(2)}(u)e_0(\ii\zeta \xi)\\
&
+\lcx_{[\alpha}P_{\beta]}\frac{m_\rho^2(e^{(\lambda)}\lcx)}{(\lcx P)^2}
\int_0^1\d u \Big\{
  \Psi^{(2)}(u)\big[e_0(\ii\zeta \xi) + 2e_1(\ii\zeta \xi) + 6e_2(\ii\zeta \xi)\big]
\nonumber\\
&\qquad\qquad\qquad
- \Psi^{(3)}(u)\big[1 +2e_0(\ii\zeta \xi) + 6e_2(\ii\zeta \xi)\big]
+ \Psi^{(4)}(u)\big[1 + e_0(\ii\zeta \xi) - 2e_1(\ii\zeta \xi)\big]
\Big\}\nonumber\\
&
+\lcx_{[\alpha}e^{(\lambda)}_{\beta]}\frac{m_\rho^2}{\lcx P}
\int_0^1\d u \Big\{
  \Psi^{(2)}(u)\big[e_0(\ii\zeta \xi)-2e_1(\ii\zeta \xi)-2e_2(\ii\zeta \xi)\big]
\nonumber\\
&\qquad\qquad\qquad\qquad
+ \Psi^{(3)}(u)\big[1+2e_2(\ii\zeta \xi)\big]
- \Psi^{(4)}(u)\big[1+e_0(\ii\zeta \xi)-2e_1(\ii\zeta \xi)\big]
\Big\}\Big] .
\nonumber
\end{align}
Now we finish the determination of the eight meson wave
functions which result from the nonlocal
light--cone quark operators of definite twist.

\section{Relations between new and conventional meson wave functions}
Obviously, since these new wave functions of mesons 
are related to true
traceless operators they differ from the conventional ones \cite{ball98,ball99}
for higher twist at least by the contributions from the trace terms.
As far as the scalar LC--operators are concerned which definitely are
of twist--2 the new and the old wave functions coincide.
However, for the vector and (skew) tensor operators also the 
contributions of dynamical twist--2 differ from those of geometric
twist--2.

Now, we are able to compare our new and the conventional meson wave functions.
We rewrite the matrix elements (\ref{O_full}), (\ref{matrix_O_nl}) and
(\ref{M_full_JJ}) by choosing
\begin{align}
\lcx_\alpha = x_\alpha -\frac{P_\alpha}{m_\rho^2} \Big(
(xP) - \sqrt{(xP)^2 - x^2 m_\rho^2}
\Big), \qquad
P_\alpha=p_\alpha+\frac{1}{2} \lcx_\alpha\frac{m_\rho^2}{\lcx P},\qquad
e^{(\lambda)}_\alpha=e^{(\lambda)}_{\bot\alpha}+p_\alpha \frac{e^{(\lambda)}\lcx}{\lcx P}-
\frac{1}{2} \lcx_\alpha m_\rho^2 \frac{e^{(\lambda)}\lcx}{(\lcx P)^2},
\end{align}
where $p_\alpha$ is a light-like vector ($p^2=0$ and $p\cdot\lcx=P\cdot\lcx$)
and $e^{(\lambda)}_{\bot\alpha}$ is the transversal polarization
vector of the $\rho$-meson. Our result is
\begin{align}
\label{JJ_full}
&\langle 0|O_{\alpha}(\kappa_1\lcx,\kappa_2\lcx)|\rho(P,\lambda)\rangle
=f_\rho m_\rho\Big[ p_\alpha\frac{e^{(\lambda)}\lcx}{\lcx P}\int_0^1\d u\, \Phi^{(2)}(u)e_0(\ii\zeta \xi)
+e^{(\lambda)}_{\perp\alpha}\int_0^1\!\!\d u
\Big\{\Phi^{(2)}(u)e_1(\ii\zeta \xi)+\Phi^{(3)}(u)
[e_0(\ii\zeta \xi)-e_1(\ii\zeta \xi)]\Big\}
\nonumber\\
&\qquad
-\frac{1}{2}\lcx_\alpha\frac{m_\rho^2(e^{(\lambda)}\lcx)}{(\lcx P)^2}\int_0^1\!\!\d u
\Big\{\Phi^{(4)}(u)\Big[e_0(\ii\zeta \xi)-3e_1(\ii\zeta \xi)
+2\int_0^1\!\!\d t\,  e_1(\ii\zeta \xi t)\Big]
-\Phi^{(2)}(u)\Big[e_1(\ii\zeta \xi)
-2\int_0^1\!\!\d t\, e_1(\ii\zeta \xi t)\Big]
\nonumber\\
&\qquad
+4\Phi^{(3)}\Big[e_1(\ii\zeta \xi)
-\int_0^1\!\!\d t\, e_1(\ii\zeta \xi t)\Big]\Big\}\Big]\ ,
\\
\label{JJ_O_nl}
&\langle 0|O^{}_{5\alpha}(\kappa_1\lcx,\kappa_2\lcx)|\rho(P,\lambda)\rangle
=
\frac{1}{2}\Big(f_\rho -f_\rho^{\rm T}\frac{m_u+m_d}{m_\rho}\Big)m_\rho
\epsilon_\alpha^{\ \,\beta\mu\nu}
e^{(\lambda)}_{\perp\beta} p_\mu\lcx_\nu
\int_0^1\d u\, \Xi^{(3)}(u) e_0(\ii\zeta \xi),
\\
\label{JJM_full}
&\langle 0|M_{[\alpha\beta]}(\kappa_1\lcx,\kappa_2\lcx)|\rho(P,\lambda)\rangle
=
\ii f_\rho^{\rm T}\Big[
2e^{(\lambda)}_{\bot[\alpha}p_{\beta]}
\int_0^1\!\!\d u\, \Psi^{(2)}(u)e_0(\ii\zeta \xi)
\\
&\qquad
+2\lcx_{[\alpha}p_{\beta]}\frac{m_\rho^2(e^{(\lambda)}\lcx)}{(\lcx P)^2}
\int_0^1\!\!\d u \Big\{
 2\Psi^{(2)}(u)e_2(\ii\zeta \xi)
-\Psi^{(3)}(u)\big[e_0(\ii\zeta \xi)+2e_2(\ii\zeta \xi)\big]
\Big\}
\nonumber\\
&\qquad
-\lcx_{[\alpha}e^{(\lambda)}_{\beta]\perp}\frac{m_\rho^2}{\lcx P}
\int_0^1\!\!\d u \Big\{
2 \Psi^{(2)}(u)\big[e_1(\ii\zeta \xi)+e_2(\ii\zeta \xi)\big]
- \Psi^{(3)}(u)\big[1+2e_2(\ii\zeta \xi)\big]
+ \Psi^{(4)}(u)\big[1+e_0(\ii\zeta \xi)-2e_1(\ii\zeta \xi)\big]
\Big\}\Big] .\nonumber
\end{align}

Comparing these expressions with the meson wave functions with dynamical twist
given by Ball~{\it et al.}~\cite{ball98,ball99}
we observe that it is necessary to re-express the truncated exponentials
and perform appropriate variable transformations. After such manipulations
we obtain the following relations, which allow us to reveal the interrelations
between the different twist definitions of meson wave functions: 
(For simplicity, we give here only the integral relations of the meson
wave functions in a state with quarks which carry momentum fraction $u$.)
\begin{align}  
\phi_\|(u)&=\Phi^{(2)}(u),\\
g_\perp^{(v)}(u)&=\Phi^{(3)}(u) + \int_u^1 \frac{\d v}{v}
\Big(\Phi^{(2)}-\Phi^{(3)}\Big)\left(v\right),\\
g_3(u)&=\Phi^{(4)}(u) - \int_u^1 \frac{\d  v}{v}\Big\{
\Big(\Phi^{(2)}-4\Phi^{(3)}+3\Phi^{(4)}\Big)\left(v\right)
+ 2 \ln \Big(\frac{u}{v}\Big)
\Big(\Phi^{(2)}-2\Phi^{(3)}+\Phi^{(4)}\Big)\left(v\right)
\Big\},\\
g_\perp^{(a)}(u)&=\Xi^{(3)}(u),\\
h_\|^{(s)}(u)&=\Upsilon^{(3)}(u),\\
\phi_\perp(u)&=\Psi^{(2)}(u),\\
h^{(t)}_\|(u)&=\Psi^{(3)}(u) + 2u \int_u^1 \frac{\d v}{v^2}
\Big(\Psi^{(2)}-\Psi^{(3)}\Big)\left(v\right),\\
h_3(u)&=\Psi^{(4)}(u)+ \int_u^1 \frac{\d v}{v}\Big\{
2\Big(\Psi^{(2)}-\Psi^{(4)}\Big)\left(v\right)
-2\frac{u}{v}\Big(\Psi^{(2)}-\Psi^{(3)}\Big)\left(v\right)
-\delta\Big(\frac{u}{v}\Big)\Big(\Psi^{(3)}-\Psi^{(4)}\Big)\left(v\right)
\Big\}.
\end{align}
We observe that both decompositions coincide in the leading terms, but 
differ at higher twist. For instance, the meson wave functions 
$g_\perp^{(v)}(u)$ and $h^{(t)}_\|(u)$ with dynamical twist $t=3$ contain 
contributions with geometrical twist $\tau = 2$ and $3$. Additionally,
dynamical twist $t=4$ meson wave functions  
$g_3(u)$ and $h_3(u)$ contain 
contributions with geometrical twist $\tau = 2$, $3$ as well as $4$.

Additionally, the conventional wave functions can be written in
terms of our new functions.
The nontrivial relations are:
\begin{align}
\Phi^{(3)}(u)&=g^{(v)}_\perp(u) +\frac{1}{u} \int_u^1 \d v
\big(g^{(v)}_\perp-\phi_\|\big)\left(v\right),\\
\Phi^{(4)}(u)&=g_3(u) 
+\frac{1}{u^2} \int_u^1 \d v\, v
\big(3g_3-4g_\perp^{(v)}+\phi_\|\big)\left(v\right)
+\frac{1}{u^2} \int_u^1 \d v\, v \Big(1-\frac{u}{v}\Big)
\big(g_3-4g_\perp^{(v)}+3\phi_\|\big)\left(v\right),\\
\Psi^{(3)}(u)&=h_\|^{(t)}(u) + \frac{2}{u}\int_u^1 \d v
\big(h_\|^{(t)}-\phi_\perp\big)\left(v\right), \\
\Psi^{(4)}(u)&=h_3(u) 
+\frac{2}{u^2} \int_u^1 \d v\, v
\big(h_3-h_\|^{(t)}\big)\left(v\right)
-\frac{2}{u^2} \int_u^1 \d v\, v \Big(1-\frac{u}{v}\Big)
\big(h^{(t)}_\|-\phi_\perp\big)\left(v\right).
\end{align}

The relation between the moments may be read off from Eqs.~(\ref{JJ_full})
-- (\ref{JJM_full}) as follows:
\begin{align}  
\phi_{\|n}&=\Phi^{(2)}_n,\\
g^{(v)}_{\perp n}&=\Phi^{(3)}_n +  \frac{1}{n+1}
\Big(\Phi^{(2)}_n-\Phi^{(3)}_n\Big),\\
g_{3n}&=\Phi^{(4)}_n -  \frac{1}{n+1}
\Big(\Phi^{(2)}_n-4\Phi^{(3)}_n+3\Phi^{(4)}_n\Big)
+\frac{2}{(n+1)^2}
\Big(\Phi^{(2)}_n-2\Phi^{(3)}_n+\Phi^{(4)}_n\Big),\\
g_{\perp n}^{(a)}&=\Xi^{(3)}_n,\\
h_{\|n}^{(s)}&=\Upsilon^{(3)}_n,\\
\phi_{\perp n}&=\Psi^{(2)}_n,\\
h_{\|n}^{(t)}&=\Psi^{(3)}_n +  \frac{2}{n+2}
\Big(\Psi^{(2)}_n-\Psi^{(3)}_n\Big),\\
h_{3n}&=\Psi^{(4)}_n+  \frac{2}{n+1}
\Big(\Psi^{(2)}_n-\Psi^{(4)}_n\Big)
-\frac{2}{n+2}\Big(\Psi^{(2)}_n-\Psi^{(3)}_n\Big)
-\delta_{n0}\Big(\Psi^{(3)}_n-\Psi^{(4)}_n\Big).
\end{align}
In terms of the moments the relations between old and new wave
functions may be easily inverted; for the wave functions itself
the expression of the new wave functions through the old ones is more
involved. The inverse relations are:
\begin{align}  
\Phi^{(3)}_{n}&=g_{\perp n}^{(v)} +  \frac{1}{n}
\Big(g_{\perp n}^{(v)}-\phi_{\|n}\Big),\qquad n>0\\
\Phi^{(4)}_{n}&=
 g_{3n}+  \frac{1}{n-1}
\Big(3g_{3n}-4g_{\perp n}^{(v)}+\phi_{\| n}\Big)
+\frac{1}{n(n-1)}\Big(g_{3n}-4g_{\perp n}^{(v)}+3\phi_{\|n}\Big),\qquad  n>1\\
\Psi^{(3)}_{n}&=h_{\|n}^{(t)}+  \frac{2}{n}
\Big(h_{\|n}^{(t)}-\phi_{\perp n}\Big),\qquad  n>0\\
\Psi^{(4)}_n&=
h_{3n}+  \frac{2}{n-1}
\Big(h_{3n}-h_{\|n}^{(t)}\Big)
-\frac{2}{n(n-1)}\Big(h_{\| n}^{(t)}-\phi_{\perp n}\Big),\qquad  n>1.
\end{align}

\section{Conclusions}
In this letter, we have discussed the calculation of the vacuum-to-meson matrix elements
for nonlocal LC-operators using the notion of geometric twist.
We have found eight meson wave functions with geometric twist $\tau$. 
From the field theoretical point of view this Lorentz invariant classification is the most appropriate frame of introducing
wave functions since the separation of different (geometric) twist 
is unique and independent from the special kinematics of the process.
An important result of our calculations are the relations between the new
meson wave functions and those given by Ball {\it et al.}~\cite{ball98,ball99}. 
These integral relations reveal 
the connection between the geometric and dynamical twist definitions for 
the meson wave functions. 

An advantage in our approach was that we use operators with well-defined
twist. Therefore, we have not used operator relations for different geometric
twist in order to isolate contributions of (geometric) twist-3 and 4 
(without gluonic contributions and neglecting quark masses) 
what is used in~\cite{ball96,stoll}.

\acknowledgments
The author is grateful to B. Geyer, P.~Ball, D.~Robaschik and S.~Neumeier 
for stimulating discussions.
M.L. acknowledges the Graduate College
``Quantum field theory'' at Center for Theoretical Sciences of
Leipzig University for financial support.

\end{document}